\documentclass[aps,prb,superscriptaddress,twocolumn]{revtex4-1}
\usepackage{color}
\usepackage{graphics}
\usepackage{graphicx} % Include figure files
\usepackage{dcolumn}% Align table columns on decimal point
\usepackage{bm}% bold math

\begin{document}

%\preprint{}

\title{Magnetic structural change of  Sr$_2$IrO$_4$ upon Mn doping}


%\author{}
%\thanks{}
%\affiliation{}

\author{S.~Calder}
\email{caldersa@ornl.gov}
\affiliation{Quantum Condensed Matter Division, Oak Ridge National Laboratory, Oak Ridge, TN 37831.}

\author{G.-X.~Cao}
\affiliation{Department of Materials Science and Engineering, University of Tennessee, Knoxville, TN 37996.}
\affiliation{Materials Science and Technology Division, Oak Ridge National Laboratory, Oak Ridge, TN 37831.}

\author{M.~D.~Lumsden}
\affiliation{Quantum Condensed Matter Division, Oak Ridge National Laboratory, Oak Ridge, TN 37831.}

\author{J. W.~Kim}
\affiliation{Advanced Photon Source, Argonne National Laboratory, Argonne, IL 60439.}

\author{Z.~Gai}
\affiliation{Center for Nanophase Materials Sciences, Oak Ridge National Laboratory, Oak Ridge, TN 37830.}

\author{B.~C.~Sales}
\affiliation{Materials Science and Technology Division, Oak Ridge National Laboratory, Oak Ridge, TN 37831.}

\author{D.~Mandrus}
\affiliation{Department of Materials Science and Engineering, University of Tennessee, Knoxville, TN 37996.}
\affiliation{Materials Science and Technology Division, Oak Ridge National Laboratory, Oak Ridge, TN 37831.}
 
\author{A.~D.~Christianson}
\affiliation{Quantum Condensed Matter Division, Oak Ridge National Laboratory, Oak Ridge, TN 37831.}


%\date{\today}

\begin{abstract}
The layered $5d$ transition metal oxide Sr$_2$IrO$_4$ has been shown to host a novel $J_{\rm eff}$=$\frac{1}{2}$  Mott spin orbit insulating state with antiferromagnetic ordering, leading to comparisons with the layered cuprates. Here we study the effect of substituting Mn for Ir in single crystals of Sr$_2$Ir$_{0.9}$Mn$_{0.1}$O$_4$ through an investigation involving bulk measurements and resonant x-ray and neutron scattering. We observe a new long range magnetic structure emerge upon doping through a reordering of the spins from the basal plane to the $c$-axis with a reduced ordering temperature  compared to Sr$_2$IrO$_4$. The strong enhancement of the magnetic x-ray scattering intensity at the $L_3$ edge relative to the $L_2$ edge indicates that the $J_{\rm eff}$=$\frac{1}{2}$ state is robust and capable of hosting a variety of ground states.
\end{abstract}

%\keywords{}

\maketitle

The discovery of a novel spin-orbit coupling (SOC) driven state in Sr$_2$IrO$_4$ has focused attention on the possibility of further unique ground states emerging from the strong SOC found in $5d$ transition metal oxides (TMO)  \cite{KimScience}.  In $5d$ TMOs  the large radius of the electronic wavefunction and increased charge results in a delicate balance between reduced Coulomb interactions, increased SOC and crystal field splitting that are all of comparable strength ($\sim$1 eV).   Investigations of $5d$ systems have led to the realization of topological insulating states \cite{NaturePesin}, Weyl semi-metal behavior \cite{PhysRevB.83.205101},  the Slater metal-insulator transition \cite{NaOsO3Calder, ShiNaOsO3}, unconventional superconductivity \cite{KOs2O6SC} and potentially the Kitaev model \cite{PhysRevLett.105.027204}.

Sr$_2$IrO$_4$ was revealed  to host a $J_{\rm eff}$=$\frac{1}{2}$ Mott spin orbit insulating state \cite{KimScience}, that has subsequently been established to exist in the iridates CaIrO$_3$ \cite{Ohgushiarxiv,PhysRevB.85.235147, PhysRevB.85.020408} and Sr$_3$Ir$_2$O$_7$  \cite{PhysRevLett.109.037204, PhysRevB.85.184432, 0953-8984-24-31-312202}. The $J_{\rm eff}$=$\frac{1}{2}$ behavior arises from the $d$-manifold degeneracy, that is split by the crystalline electric field into the $t_{2g}$ and $e_{g}$ levels, being further broken by SOC driving the $t_{2g}$ level into a lower energy $J_{\rm eff}$=$\frac{3}{2}$ degenerate quadruplet and a high energy  $J_{\rm eff}$=$\frac{1}{2}$ degenerate doublet. For the $5d^5$ Ir$^{4+}$ ion this results in a filled $J_{\rm eff}$=$\frac{3}{2}$ manifold and a half filled $J_{\rm eff}$=$\frac{1}{2}$ manifold that is split by even the small on-site Coulomb interactions in $5d$ TMOs. The similarity of the magnetic insulating state and the crystal structure in Sr$_2$IrO$_4$ with the layered cuprate La$_2$CuO$_4$ has led to interest in realizing superconductivity via doping Sr$_2$IrO$_4$ \cite{PhysRevLett.108.177003, PhysRevLett.106.136402}.   Indeed an important open question is whether or not the $J_{\rm eff}$=$\frac{1}{2}$ state is robust to perturbations, such as doping, and consequently able to act as a vehicle from which unusual ground states can emerge.

Sr$_2$IrO$_4$ crystalizes into the tetragonal $I4_1/acd$ space group with the IrO$_6$ octahedra rotated around the $c$-axis by $\sim$11$^{\circ}$.  \cite{PhysRevB.49.9198, Huang1994355} This results in the reduced symmetry of $I4_1/acd$ compared to $I4/mmm$. Sr$_2$Ir$_{0.9}$Mn$_{0.1}$O$_4$, the material of interest in this work, has Mn substituted on the Ir site that causes no symmetry change, as we confirmed by x-ray diffraction (XRD). For comparison a full series powder XRD investigation of Sr$_2$Ir$_{1-x}$Ti$_x$O$_4$ reported a change from $I4_1/acd$ to $I4/mmm$ for $x$$\approx$0.4, with a similar doping value inferred for  Fe and Co\cite{Gatimu2012257}. The end member for our chosen doping, Sr$_2$MnO$_4$, indeed forms the $I4/mmm$ space group \cite{Tezuka1999705}.

We present the first study of Mn-doped Sr$_2$IrO$_4$. Resonant x-ray scattering (RXS) has emerged as an important and powerful tool in the investigation of $5d$ magnetism, revealing the $J_{\rm eff}$=$\frac{1}{2}$ state in Sr$_2$IrO$_4$ and the long range magnetic structure \cite{KimScience}. We employ this technique to test the robustness of the $J_{\rm eff}$=$\frac{1}{2}$ state in Mn-doped Sr$_2$IrO$_4$ and to observe alterations to the magnetic structure. Neutron scattering allows for a direct comparison between the observed and calculated magnetic Bragg scattering intensities. We use this technique to distinguish between symmetry allowed magnetic structures and obtain an ordered magnetic moment.  We report complimentary Sr$_2$Ir$_{0.9}$Mn$_{0.1}$O$_4$ single crystal measurements that consider changes in bulk magnetic correlations through magnetization and resistivity. 

Single crystals of Sr$_2$Ir$_{0.9}$Mn$_{0.1}$O$_4$ were grown in a Pt crucible using the flux method. Neutron scattering was performed on an 11 mg single crystal, dimensions 0.2$\times$0.2$\times$0.02cm, at the High Flux isotope reactor (HFIR) on the triple axis instrument HB-3 in elastic mode with $\lambda=2.36$ $\rm \AA$. Pyrolytic graphite (002)  monochromator and analyzer were used and the collimation set to 48'-80'-80'-240'. The sample was mounted in the ($H0L$) scattering plane. Due to the form factor decrease in intensity with scatting angle, high neutron absorption of iridium and small sample mass counting times of seconds to several minutes were employed to obtain sufficient magnetic reflections. While the flat plate geometry of the sample suppressed neutron absorption, we calculated the absorption at each Bragg peak based on the geometry of the sample and subsequent beam pathway through the volume of the sample. The absorption factor for Sr$_2$Ir$_{0.9}$Mn$_{0.1}$O$_4$ is 5.3cm$^{-1}$. RXS measurements on a crystal of approximate size 1$\times$1$\times$0.5 mm were performed at the Advanced Photon Source (APS) at beamline 6-ID-B. We carried out measurements at both the $L_2$ (12.82 keV) and $L_3$ (11.22 keV) resonant edges of iridium. Graphite was used as the polarization analyzer at the (0 0 10) and (008) reflections on the $L_2$ and $L_3$ edges, respectively, to achieve a scattering angle close to 90$^\circ$. Measurements were taken at several reflections to investigate possible magnetic structures, with an analysis of the photon polarization in $\sigma$-$\sigma$ and $\sigma$-$\pi$ allowing magnetic and charge scattering  to be distinguished. To observe the sample fluorescence, energy scans were performed without the analyzer and with the detector away from any Bragg peaks through both absorption energies. The sample magnetization $M(T,H)$ was measured with a Quantum Design (QD) magnetic property measurement system (MPMS) in applied fields up to 7 T. The electrical resistivity  $\rho(T,H)$ were performed using a QD 14T physical property measurement system (PPMS). 
 
%trim is left, bottom, right, top
\begin{figure}[tb]
     \centering   
\includegraphics[width=1.0\columnwidth]{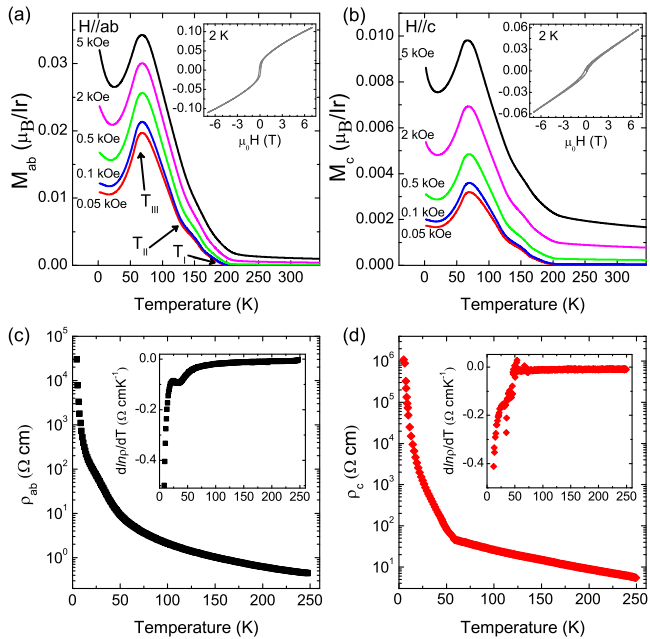}
 \caption{\label{FigureBulk}  Field cooled magnetization measurements on single crystals of Sr$_2$Ir$_{0.9}$Mn$_{0.1}$O$_4$ for applied fields  parallel to the (a) $ab$-plane and (b) $c$-axis. Inset shows the isothermal magnetization from -7 to 7 T at 2 K. Zero field resistivity measurements along the (c) $ab$-plane and (d) $c$-axis.}
\end{figure}

Undoped Sr$_2$IrO$_4$ forms a long-range AFM structure at 240 K with a net ferromagnetic moment arising from canting of the spins within the basal plane, with the degree of canting governed by the angle of rotation of the octahedra \cite{PhysRevLett.102.017205, KimScience}.  Magnetization results for Sr$_2$Ir$_{0.9}$Mn$_{0.1}$O$_4$ (Fig.~\ref{FigureBulk}(a)-(b)) show the onset  of magnetic correlations occur around $\sim$170 K for low field  cooled measurements, significantly reduced compared to Sr$_2$IrO$_4$. Increasing the applied field causes an increase in the transition temperature. The value of $0.11$$\mu_B$ at 2 K in the maximum field applied of 7 T (see Fig.~\ref{FigureBulk}(a) inset) is close to the saturated moment in the literature for Sr$_2$IrO$_4$ of  0.14$\mu_B$ \cite{PhysRevB.57.R11039}.  The magnetization anisotropy between the $c$-axis and the $ab$-plane persists upon Mn-dopoing, with the magnetization remaining largest for a field in $ab$-plane. Discrete regions, separated by anomalies in the magnetization and resistivity, were evident in measurements for Sr$_2$IrO$_4$ \cite{GeSr2IrO4}. A similar distinction can be applied to Sr$_2$Ir$_{0.9}$Mn$_{0.1}$O$_4$ with anomalies in the magnetization at  $\rm T_{I}$=$170$ K,  $\rm T_{II}$=$125$ K and  $\rm T_{III}$$\sim$$55$ K for the 0.05 kOe results, highlighted in Fig.~\ref{FigureBulk}(a). Increasing the applied field leads to a removal of the $\rm T_{II}$ anomaly. For Sr$_2$IrO$_4$ at low temperature ($<$20 K) the magnetization increased along the $c$-axis and decreased along the $ab$-plane, leading to the suggestion of increased canting of spins along the $c$-axis.\cite{GeSr2IrO4} Figure \ref{FigureBulk}(a)-(b), however, shows for Sr$_2$Ir$_{0.9}$Mn$_{0.1}$O$_4$ that M$_{ab}$ and M$_c$ both increase, indicating no enhanced $c$-axis canting.
 
In Sr$_2$IrO$_4$ the transition from a high temperature metallic phase to low temperature insulating state occurs without a well defined boundary above room temperature, with debate as to the possibility of a combination of Mott and magnetic correlations via the Slater mechanism driving the insulating state \cite{AritaSlater}. For Sr$_2$Ir$_{0.9}$Mn$_{0.1}$O$_4$  the resistivity remains unchanged through both $\rm T_{I}$ and  $\rm T_{II}$, see Fig.~\ref{FigureBulk}(c)-(d). The lack of a discernible change in the resistivity  at the upper magnetic transitions indicates that the initial onset of magnetic order does not affect the electron scattering rate and therefore spin disorder scattering is not a significant term in the resistivity in this region. At $\rm T_{III}$, however,  concurrent with the sharp drop in the magnetization, the resistivity increases at a greater rate. The derivative of the resistivity shows this change to occur more clearly for both the $c$-axis and $ab$-plane, shown inset in Fig.~\ref{FigureBulk}(c)-(d). Therefore the $\rm T_{III}$ region shows a coupling of the magnetization and resistivity in three-dimensions, potentially due to a structural distortion or a change in the magnetic correlations that localizes the conduction electrons.

\begin{figure}[tb]
     \centering                      
\includegraphics[trim=0.05cm 0.32cm 0.1cm 0.1cm, clip = true, width=0.88\columnwidth]{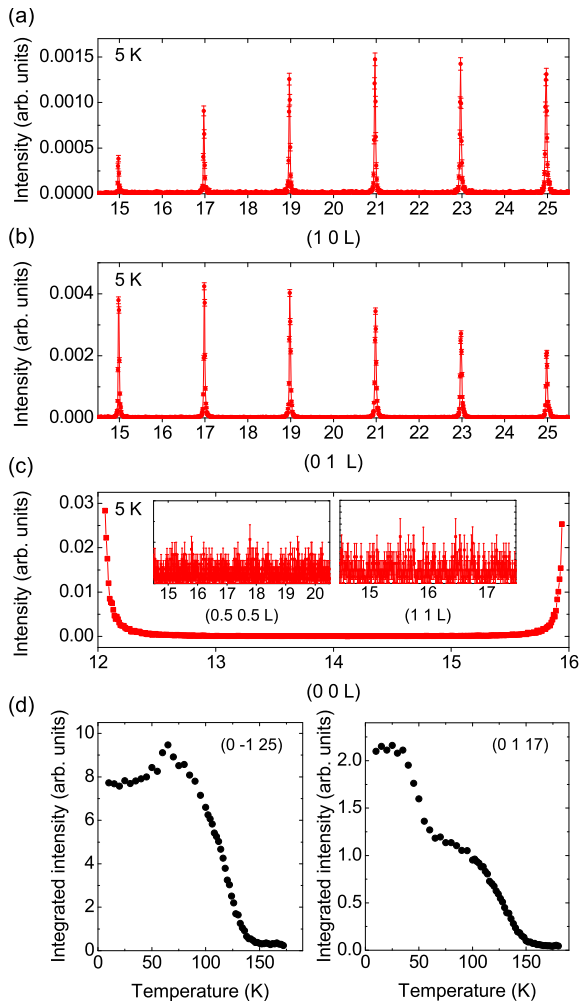}
 \caption{\label{FigureLongLscans} Elastic RXS of Sr$_2$Ir$_{0.9}$Mn$_{0.1}$O$_4$ at the Ir $L_3$-edge in $\sigma$-$\pi$ mode. (a)-(b) L scans at constant (H,K) reveal long range magnetic order  at (1 0 $odd$) and (0 1 $odd$) at 5 K. The same reflections are present at 80 and 120 K. (c) No magnetic reflections are observed at (0 0 L), (1 1 L) or (0.5 0.5 L) positions at 5, 80, 120, 160 or 170 K. (d) The change in integrated intensity with temperature for two magnetic reflections.}
\end{figure}

Having established 10$\%$ Mn-doping of Sr$_2$IrO$_4$ produces alterations in the magnetic correlations we consider the long range magnetic ordering through a combination of RXS and neutron scattering. RXS results, presented in Fig.~\ref{FigureLongLscans}(a)-(c), show magnetic order for Ir ions is characterized by (1 0 $odd$) and (0 1 $odd$) reflections in Sr$_2$Ir$_{0.9}$Mn$_{0.1}$O$_4$. No magnetic scattering is observed at  (00L), (11L), ($\frac{1}{2}$$\frac{1}{2}$L) reflections or at incommensurate positions at 5, 80, 120, 160 or 170 K for the regions shown in Fig.~\ref{FigureLongLscans}(c). We note that the large charge scattering in RXS will obscure any potential magnetic scattering at (0 0 4$n$). 

%trim is left, bottom, right, top

Figure \ref{FigureLongLscans}(d) shows the temperature dependence of the integrated intensity for select magnetic peaks. The onset of long range magnetic order occurs at $\sim$155 K. This is reduced compared to the applied-field magnetization results. This is, however, consistent with the observation from magnetization measurements that the application of even small fields causes an increase in the magnetic transition temperature in Sr$_2$Ir$_{0.9}$Mn$_{0.1}$O$_4$ and as such the zero field RXS should produce a lower $\rm T_N$. To test for x-ray beam heating we compared the temperature dependence for an attenuator in the beam providing approximately a factor of 2 attenuation and no difference in the ordering temperature was observed.  We cannot additionally rule out poor thermal contact or sample variation compounding this effect.   Despite both chosen reflections showing the same magnetic ordering temperature, the lower temperature region around $\rm T_{III}$ showed anomalous behavior. To search for potential changes in the long range magnetic structure through this region we performed the same scans presented in Fig.~\ref{FigureLongLscans}(a)-(c) at various temperatures between 5 K and 170 K and observed no removal or development in the measured magnetic Bragg peaks. Therefore, although the long range magnetic structure remains unaltered, the region around $\rm T_{III}$ appears to host a change consistent with changes in the moment direction or a small structural distortion.

Comparing our results with RXS  measurements on Sr$_2$IrO$_4$ demonstrates Mn doping  forces an alteration to a new magnetic structure, as evidenced by the difference in the magnetic Bragg positions due to a change in the ordering wavevector. Sr$_2$IrO$_4$ has magnetic scattering at (0 0 $odd$) as well as (1 0 4$n$+2) and (0 1 4$n$) reflections. The observation in  Sr$_2$IrO$_4$ that  (1 0 L) $\ne$ (0 1 L) requires a symmetry transition from tetragonal to orthorhombic \cite{KimScience}. Any potential departure from tetragonal is not observed within the resolution of our measurements for Sr$_2$Ir$_{0.9}$Mn$_{0.1}$O$_4$. The same study of Sr$_2$IrO$_4$ found an application of a small in-plane field of H $>$ 0.2 T altered the magnetic structure such that the (1 0 $4n$$+$2) peaks disappear and new peaks appear at (1 0 $odd$) reflections. This allows us to speculate that Mn-doping of Sr$_2$IrO$_4$ simulates the behavior of Sr$_2$IrO$_4$ in a small magnetic field. However, as we show below, the in-field magnetic structure presented by Kim {\it et al.} \cite{KimScience} from RXS is not compatible with the neutron scattering intensities observed for Sr$_2$Ir$_{0.9}$Mn$_{0.1}$O$_4$.

\begin{figure}[tb]
     \centering   
\includegraphics[width=0.49\columnwidth]{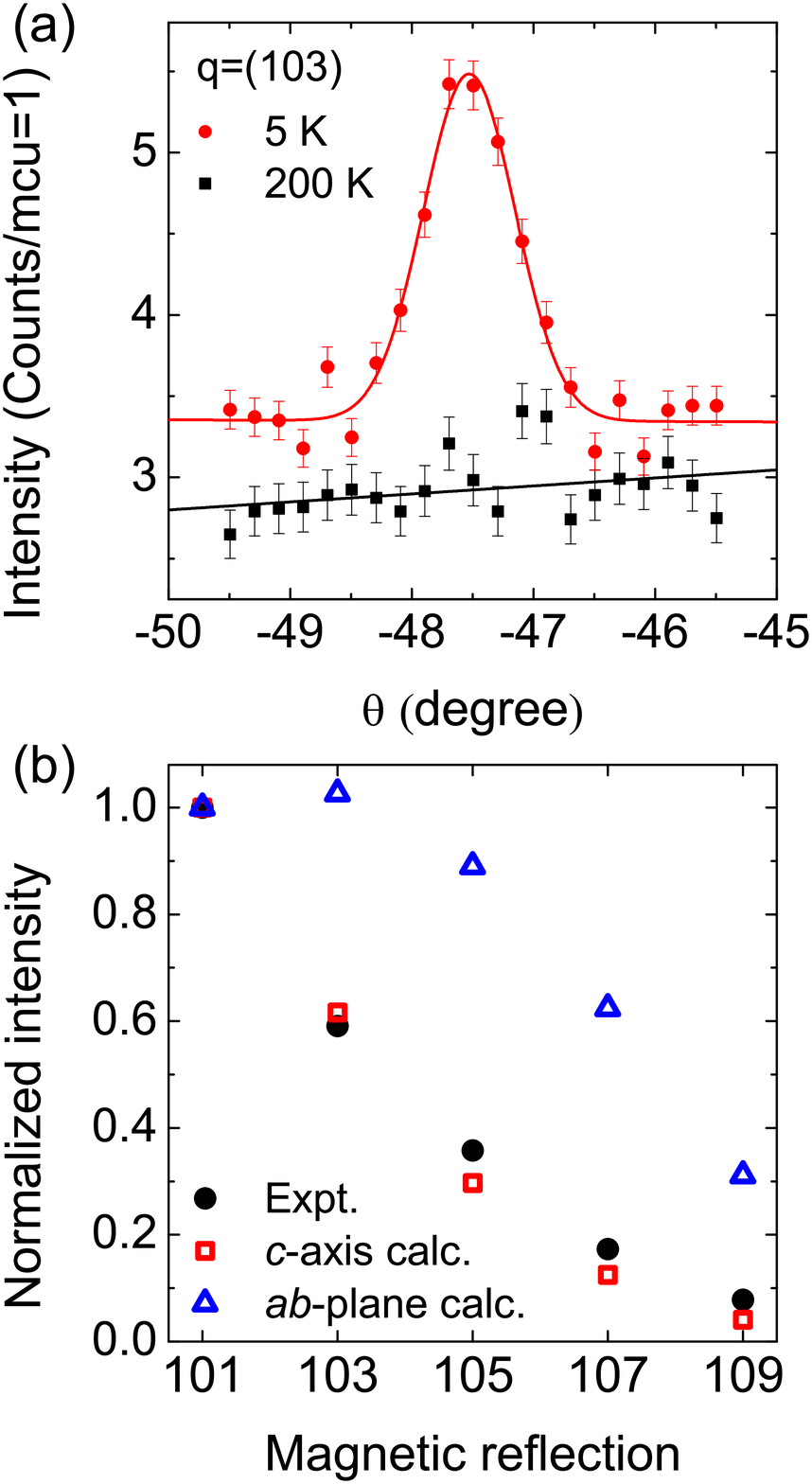}
\includegraphics[width=0.41\columnwidth]{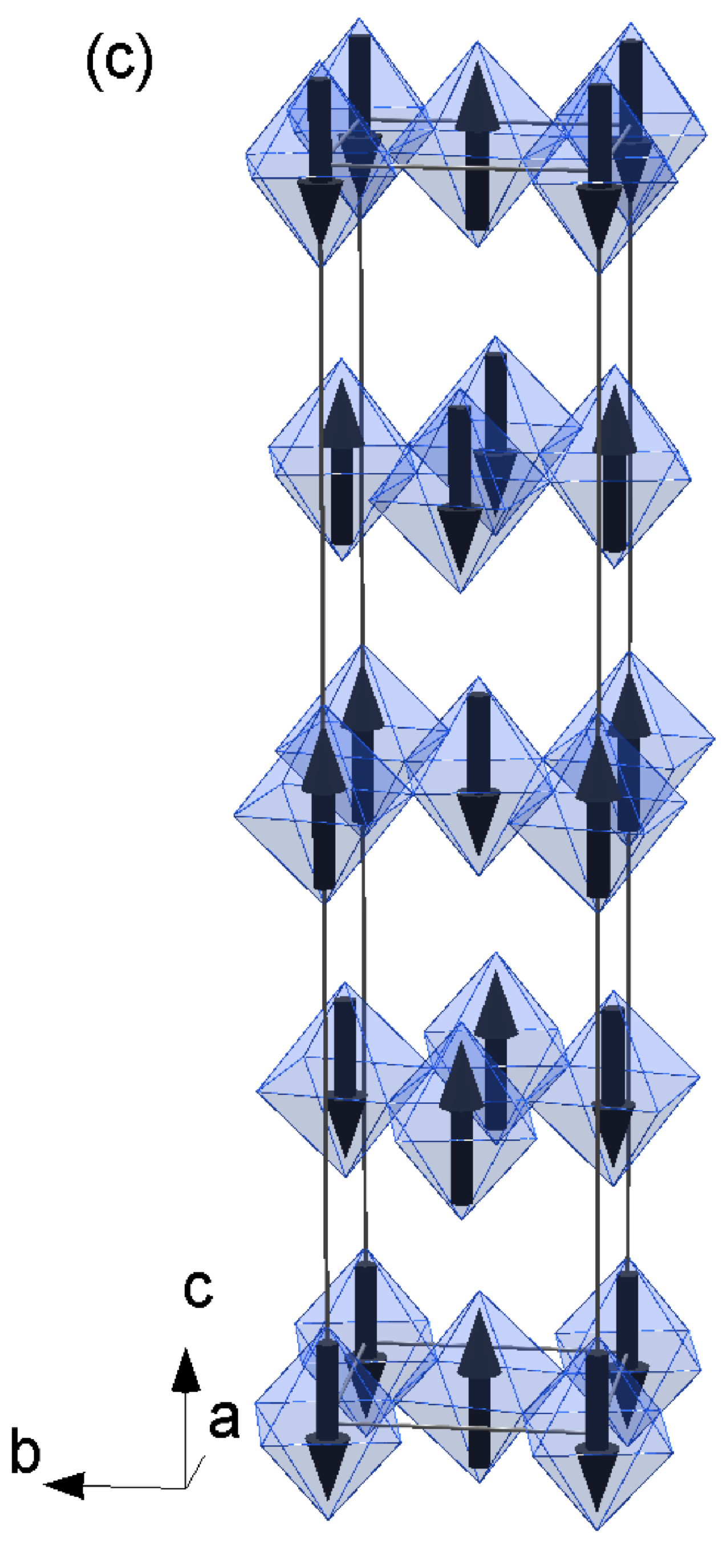}
 \caption{\label{Figuremagneutron}  Single crystal neutron scattering measurements.~(a) Rocking scans of the (103) magnetic reflection above and below the magnetic transition temperature. Similar scans were performed at the (101),(105),(107) and (109) magnetic reflections.~(b) The normalized experimental intensity for the measured reflections is compared to instrument resolution and absorption corrected calculated scattering for both magnetic moments in the $ab$-plane and along the $c$-axis.~(c) The long ranged magnetic structure for Sr$_2$Ir$_{0.9}$Mn$_{0.1}$O$_4$.}
\end{figure}

The unpredictable effects of multiple scattering and the variable sample absorption from our RXS measurements made it unfeasible to obtain quantitative information regarding the new long range magnetic structure from either a peak intensity comparison at different reflections or from an azimuthal scan. Instead we utilized the scattering values obtainable from single crystal neutron scattering. Such results are presented in Fig.~\ref{Figuremagneutron}. Measurements were performed at (10$L$) reflections for $L$=1,3,5,7,9 and experimental peak intensities extracted. Calculated intensities were obtained for different magnetic models with the inclusion of the instrument resolution at each scattering angle using ResLib \cite{ResLib}, absorption correction using the sample and scattering geometry at each reflection and magnetic form factor for Ir$^{4+}$.\cite{Kobayashiformfactor}

%trim=4.0cm 0.35cm 2.5cm 0.15cm, 
%trim is left, bottom, right, top

To explore the specific nature of the long range magnetic ordered structure we implemented representational analysis \cite{sarahwills}. The propagation vector ${\bf k}$$=$(000) was employed in our analysis as it is consistent with all the observed magnetic reflections.  For a second order transition, Landau theory states that the symmetry properties of the magnetic structure are described by only one irreducible representation (IR). The space group $I4_1/acd$ and Ir ions on the $8a$ wyckoff position gives the IRs: $\Gamma_1$, $\Gamma_3$, $\Gamma_6$, $\Gamma_8$, $\Gamma_9$ and $\Gamma_{10}$ (following the numbering scheme of Kovalev  \cite{Kovalev}).  $\Gamma_9$ and $\Gamma_{10}$ describe magnetic structures with spins in the basal plane, as has been presented for the parent Sr$_2$IrO$_4$ compound  \cite{KimScience}. Magnetic reflections at (1 0 $odd$) and (0 1 $odd$) can be generated using these IRs. The experimental intensity at the magnetic Bragg peaks are compared with the calculated intensity for Sr$_2$Ir$_{0.9}$Mn$_{0.1}$O$_4$ in Fig.~\ref{Figuremagneutron}(b). Poor agreement is found, indicating a magnetic structure with the spins in the $ab$-plane is not compatible with the measured scattering. $\Gamma_3$  gives a purely ferromagnetic magnetic structure and therefore the observed scattering is not reproduced. $\Gamma_6$ produces antiferromagnetic spins along the $c$-axis and ferromagnetic interactions in each layer, this structure does not reproduce the observed scattering at (1 0 $odd$) or (0 1 $odd$).  The only remaining IRs,  $\Gamma_1$ and $\Gamma_{8}$, have antiferromagnetic spins along the $c$-axis. Both models give identical scattering intensity and symmetry equivalent magnetic structures through a transformation of $a$$\rightarrow$$b$. The calculated magnetic scattering at (1 0 $odd$) reflections is shown in Fig.~\ref{Figuremagneutron}(b), with close agreement between experimental and calculated intensities for all the reflections measured. The corresponding magnetic structure is shown in Fig.~\ref{Figuremagneutron}(c). Considering the ratio between several nuclear and magnetic peaks yields an ordered magnetic moment of 0.5(1)$\mu_B$/Ir. Thus we can conclude that the substitution of 10$\%$ Mn on the Ir site in Sr$_2$IrO$_4$ results in a flipping of the spins from the basal plane to the $c$-axis. The scattering for Mn-doped Sr$_2$IrO$_4$ occurs at the same reflections as that for Sr$_2$IrO$_4$ in a small applied field of 0.2 T \cite{KimScience}. The magnetic structure presented for a small applied field is identical to that for Sr$_2$Ir$_{0.9}$Mn$_{0.1}$O$_4$, with the spins flipped from the $ab$ to c-axis. A similar spin-flop transition was shown to occur in going from Sr$_2$IrO$_4$ ($ab$-plane) to Sr$_3$Ir$_2$O$_7$ ($c$-axis) due to a change in dimensionality \cite{PhysRevLett.109.037204}. This suggests that the magnetic ordered state in Sr$_2$IrO$_4$ is potentially unstable and can be influenced by other perturbations.

\begin{figure}[t] 
     \centering                      
\includegraphics[trim=1.5cm 0.35cm 0.35cm 0.15cm, clip = true, width=0.85\columnwidth]{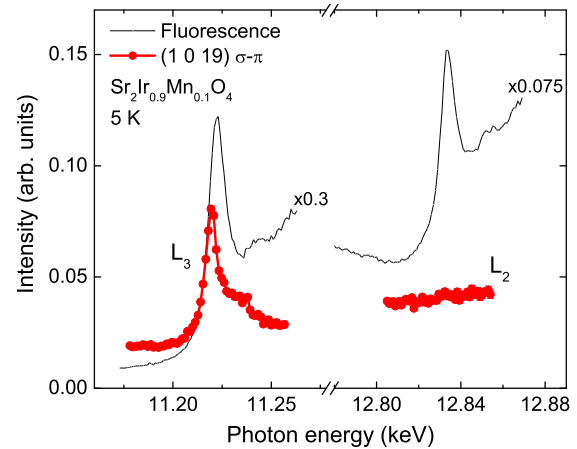}
 \caption{\label{FigureMRXSLedges} RXS energy dependence of the iridium L$_3$-edge at 11.22 keV and the L$_2$-edge at 12.82 keV.  These correspond to the inflection point at the low energy side of the enhancement in the fluorescence scans. Measurements at both edges of Sr$_2$Ir$_{0.9}$Mn$_{0.1}$O$_4$ only yields appreciable resonant enhancement at the L$_3$ edge for $\sigma$-$\pi$ magnetic scattering. This behavior is a signature of $J_{\rm eff}$=$\frac{1}{2}$ state.}
\end{figure}

The resonant enhancement for Ir not only allows RXS to be utilized to observe long range magnetic ordering, but gives a signature of the $J_{\rm eff}$=$\frac{1}{2}$ Mott insulating state \cite{KimScience}. In a non-SOC split Ir$^{4+}$ $t_{2g}$ manifold a large enhancement is expected at both $L_2$ and $L_3$ edges, however  the scattering at $L_2$ is forbidden for the $J_{\rm eff}$=$\frac{1}{2}$ SOC induced state. Therefore a measurement of the intensity through the resonant energies at magnetic reflections provides direct evidence for the existence of the $J_{\rm eff}$=$\frac{1}{2}$ state. Such measurements are presented in Fig.~\ref{FigureMRXSLedges} at the (1 0 19) magnetic reflection. A large enhancement is observed at the $L_3$ edge that has its maximum at the inflection point of the fluorescence scans, as expected. Conversely there is no appreciable resonant enhancement at the $L_2$ energy. The $J_{\rm eff}$=$\frac{1}{2}$ state remains in Sr$_2$Ir$_{0.9}$Mn$_{0.1}$O$_4$, despite the change in magnetic structure.   The $J_{\rm eff}$=$\frac{1}{2}$  state has been shown to exist in iridates other than Sr$_2$IrO$_4$, that host alternative crystal symmetries \cite{PhysRevB.85.235147,PhysRevLett.109.037204, PhysRevB.85.184432}, and now upon diluting the Ir site with a $3d$ magnetic ion that has reduced SOC and cannot itself host the state. This points to the robustness of the $J_{\rm eff}$=$\frac{1}{2}$ state despite the apparent ease of altering the magnetic structure it inhabits. It would be of interest to follow Mn-doped Sr$_2$IrO$_4$, or an alternative doping,  through the expected $I4_1/acd$ to $I4/mmm$ doping driven structural  transition to follow any further change in long-range ordering and the subsequent effect on the $J_{\rm eff}$=$\frac{1}{2}$ state.

%trim is left, bottom, right, top

Our results show that substituting Mn for Ir in Sr$_2$IrO$_4$ leads to an alteration of the magnetic structure through a reordering and flipping of the spins.  The effect of 10$\%$ Mn doping produces magnetic Bragg positions consistent with the application of a small field of 0.2 T to Sr$_2$IrO$_4$, however the presented magnetic structures differ through a flipping of each spin from the $ab$ plane to $c$-axis. The onset of magnetic ordering is reduced from 240 K to $\sim$155 K and controllable by applied fields. Even with the altered magnetic ordering of the Ir ions the $J_{\rm eff}$=$\frac{1}{2}$  insulating state remains.  Despite speculation as to the possibilities of doping Sr$_2$IrO$_4$, few experimental studies exist. The results presented suggest an unstable magnetic structure in Sr$_2$IrO$_4$ that can be altered by small perturbations, whereas breaking the $J_{\rm eff}$=$\frac{1}{2}$ state appears to require a more dramatic alteration. Further studies using alternative doping ions and concentrations will lead to a greater level of understanding in this important $5d$ material.  

%\begin{acknowledgments}
We thank Ling Li for supporting characterization measurements. This research at ORNL's High Flux Isotope Reactor was sponsored by the Scientific User Facilities Division, Office of Basic Energy Sciences, U.S. Department of Energy. Part of the work (DM, BCS, GC) was supported by the Department of Energy, Basic Energy Sciences, Materials Sciences and Engineering Division. A portion of this research was conducted at the Center for Nanophase Materials Sciences, which is sponsored at Oak Ridge National Laboratory by the Scientific User Facilities Division, Office of Basic Energy Sciences, U.S. Department of Energy. Use of the Advanced Photon Source, an Office of Science User Facility operated for the U.S. DOE Office of Science by Argonne National Laboratory, was supported by the U.S. DOE under Contract No. DE-AC02-06CH11357.
%\end{acknowledgments}

 %\bibliography{MndopedSr2IrO4}

%merlin.mbs 2010-03-15 4.21a (PWD, AO, DPC)

\end{document}